# Designing Culturally Aware Learning Analytics: A Value Sensitive Perspective

*Olga Viberg*
KTH Royal Institute of Technology
Lindstedsvägen 3, 10044 Stockholm, Sweden
oviberg@kth.se

*Ioana Jivet*
DIPF Leibniz Institute for Research and Information in Education
Rostocker Straße 6, 60323 Frankfurt am Main, Germany
jivet@dipf.se

*Maren Scheffel*
Ruhr-Universität Bochum
Institut für Erziehungswissenschaft
Universitätsstraße 150, D - 44801 Bochum, Germany
e-mail: maren.scheffel@rub.de

**Abstract**
This chapter aims to stress the importance of addressing culture when designing and implementing learning analytics services. Learning analytics have been implemented in different countries with the purpose of improving learning and supporting teaching; yet, largely at a limited scale and so far with limited evidence of achieving their purpose. Even though some solutions seem promising, their transfer from one country to another might prove challenging and sometimes impossible due to various technical, social, contextual and cultural factors. In this chapter, we argue for a need to carefully consider one of these – largely underexplored by the learning analytics community – factors, namely cultural values when designing and implementing learning analytics systems. Viewing culture from a value-sensitive perspective, in this chapter, we: 1. exemplify two selected values (i.e., privacy and autonomy) that might play a significant role in the design of learning analytics systems, and 2. discuss opportunities for applying culture- and value-sensitive design methods that can guide the design of culturally aware learning analytics systems. Finally, a set of design implications for *culturally aware* and *value-sensitive* learning analytics services is offered.

## 10.1 Introduction

Learning analytics (LA) has been implemented and used in various countries in different ways, often at a limited scale (Viberg et al. 2018).

Across countries and continents, there are differences in the expectations teachers and students have towards LA (Hilliger et al. 2020; Kollom et al. 2021; Pontual Falcao et al. 2022; Viberg et al. 2022), as well as different concerns about the ethical issues surrounding LA (West et al. 2018; Hoel and Chen 2019; Mutimukwe et al. 2022). These differences make the transfer of LA solutions from one country to another challenging, i.e., varying contextual, technical, and also cultural factors may play an important role. Whereas technical and contextual aspects of LA systems' design and implementation have been addressed by LA scholars and practitioners, cultural factors have so far received scarce attention (Jivet et al. 2022). Paying attention to culture – both at the individual and also, at the national level – might be an endeavour worth exploring. As we argue later in this chapter, various cultural factors may influence students' or teachers' behavioural intentions and their eagerness to accept and adopt new technologies (Viberg and Grönlund 2013). For example, already in 1962, Rogers (1962) investigated various factors behind individuals' different levels of adoption of new technologies. He suggested dividing users according to their time to adoption (e.g., from the early innovators to early adopters). Scholars have studied different factors behind this time distribution, including gender, age, and technical skills, and suggested that many of them can have more than a temporal character. However, cultural factors have been argued to be more important since they are supposed to have more longevity, transcending the development of individuals (Viberg and Grönlund 2013). This would make cultural factors highly interesting to consider and investigate in the LA setting even though they are often perceived to be challenging to study directly. As an alternative, they are often studied through some proxies, such as cultural values (e.g. Hofstede et al. 2010; Milberg et al. 2000).

The idea that a 'one size fits all' paradigm does not lead to effective LA tool designs and implementation has been accepted within both the technology-enhanced learning and the LA communities (Gašević et al. 2016; Teasley 2017; Jivet 2021). However, there is still a question about what factors define the 'right size', and throughout this chapter, we make a case for considering culture as one of these factors.

In this chapter, we argue that culture might play a role in the design of LA and discuss possible cultural differences – the factors that have so far not been extensively studied by LA researchers – for the wider successful adoption of LA at scale. In particular, this chapter discusses whether the stakeholders' (e.g., students' and teachers') *cultural values* are some of these factors. In an increasingly international educational landscape, how and to what extent should the LA community take into account such factors in order to have a significant impact (i.e., to improve learning and teaching) at scale? What opportunities are offered by LA technologies to consider stakeholders' cultural preferences and values, and how can we design culturally aware LA services which account for these values?

**10.2 Why is culture relevant for LA?**

In general, a careful understanding of culture is important to the study of information technologies. Culture at various levels, including national, organisational, and groups, can influence the successful implementation and use of information technology (Leidner and Kayworth 2006; Lee et al. 2013). This understanding is similarly critical to the successful implementation of LA systems to ensure equal and fair learning support opportunities for students from diverse cultures and in different educational settings. Cultural pluralism can lead to positive learning outcomes, including improved interaction skills, working relationships, and improved cognitive reasoning (Johnson and Jonson 1989). At the same time, when not addressed appropriately, "cultural diversity in [collaborative] learning can lead to negative relationships characterised by hostility, rejection, stereotyping, and prejudice" (Economides 2008, p. 249). Thus, we argue that culture and cultural differences should be considered in the design and implementation of LA solutions to both enhance learning and minimise adverse effects of culturally diverse learning environments.

While there have been some initial attempts to focus LA on cultural differences (Vatrapu 2011), this critical topic is largely under-researched in current LA research and practice (Jivet et al. 2022). The research efforts on the topic point out at least three research directions. *First,* there are cultural differences in learning and teaching which will inevitably shape any LA used to analyse those processes, but also tools developed for these settings. Marambe et al. (2012) have, for example, shown that student learning patterns and learning strategies in higher education differ across cultures. Cultural differences were also shown to play a role in online learning settings as they influence students' collaborative learning (Vatrapu and Suthers 2007), which is in line with broader research on culturally-aware collaborative learning (Economides 2008) and self-regulation (Lin et al. 2021; McInerney 2008; Purdie & Hattie 1996).

*Second,* there are cultural differences in responses to LA related to, for example, adoption and the effectiveness of interventions. Nistor et al. (2013) showed that educational technology acceptance is influenced by culture with members of masculine cultures (cf. Hofstede 2001) primarily expecting educational technology to improve their learning performance and members of individualistic cultures (cf. Hofstede 2001) being less susceptible to social influence. Furthermore, Mittelmeier et al. (2016) found that cultural diversity explained a substantial part of the variation in learning dispositions, like boredom and learning enjoyment, as well as the use intensity of e-tutorials as part of a blended learning course. When analysing the effectiveness of interventions in the context of massive open online courses (MOOCs), Kizilcec & Cohen (2017) showed that a writing activity that facilitates goal-commitment and goal-directed behaviour raised educational attainment at scale in individualist (US and Germany) but not in collectivist cultures such as China. Later, Cho et al. (2021) examined the degree to which social norm messages can motivate people in different countries to persist in online learning and engage in their peer community and found that both the type of norm message (e.g., descriptive or injunctive) and the cultural context (China, US) influenced how the intervention improved course outcomes. Finally, Davis et al. (2017) showed that when learners are exposed to a learning dashboard that facilitates social comparison learners from countries with weak social norms and high tolerance for deviant behaviour significantly outperform their peers from countries with strong social norms and a low tolerance for deviant behaviour in terms of both engagement and achievement.

*Third,* LA can be used to study cultural differences in learning and teaching, especially in educational scenarios with culturally diverse populations. For example, Ruipérez et al. (2022) found evidence that MOOC learners feel more comfortable and at ease when learning in their native language and having instructors with a similar cultural background, while Rizvi et al. (2022) have recently shown that certain types of learning activities (e.g., discussion) facilitate the progress of Anglo-Saxon learners while inhibiting the progress of learners from South Asia.

Overall, these examples show the importance of designing culturally aware or culture-sensitive (these two terms are used interchangeably in the present chapter) LA services that, as we argue, would increase the acceptance and adoption of LA at a global scale.

*Cultural sensitivity* is explained as "the competence to be aware of and to experience differences and similarities between people – their values and practices – and that they are based on what they have learned as members of groups" (Van Boeijen and Zijlstra 2020, p.20). Following this, the goal of the culture-sensitive LA designer is to know what the values, needs, and desires of the intended users (students and teachers) are, grounded in who they are as a part of the cultural group. Based on the earlier design-oriented research efforts in other fields (e.g., human-computer interaction), scholars suggest five possible intentions that offer a direction and are not a fixed outcome – as compared to aims or strategies – to be considered in dealing with culture (Van Boeijen and Zijlstra 2020). They are: **1**. to affirm a culture; **2.** to attune to a culture; **3**. to change culture; **4.** to abridge cultures, and **5**. to bypass cultures.

*To affirm a culture* in the LA setting suggests that the designer's intention is to acknowledge and endorse the existing cultural values of the targeted users. Such values may include the stakeholders' values of for example, collaboration, privacy, trust, transparency, and openness, which are all important to LA system development and an LA services' further acceptance. For example, in strongly individualistic cultures like the US culture (Hofstede 2001), this aspect could be upheld by way of LA services for highly individual and personalised usage. Another example refers to the different levels of individuals' trust. In this regard, for example, the Nordic nations as compared to the originally more heterogeneous cultures such as the US, share a unique bond through cultural identity, creating an environment where these societies can exist with high levels of trust, transparency, and openness (Robinson 2020). Thus, the goal of the LA designer will be to affirm these values when designing LA services, and at the same time, to 'protect' users by undertaking a 'responsible' design approach that would enable their agency.

*Attuning to a culture* suggests that the designer intentionally "focuses on the attempt to be in tune with existing cultural values in order to achieve an optimal design and to avoid mismatches between the cultural group and the product" (Van Boeijen and Zijlstra 2020, p. 25). In this, the LA designer may need to consider: *forms, colours, symbols, properties* (that describe the expected behaviour of the LA service under certain circumstances), *functions* (e.g., what the user can do with it; this can be specific for a cultural group), *interactions* (e.g., people have learnt to interact in some certain ways in some cultures, and in other ways in some other cultures; these interactions may be difficult to change), *needs*, and *values* (i.e., how people value a service or a tool is affected by the cultural context in which they have learnt what is morally right or wrong, or good or bad). In general, the design of a LA service needs to be attuned to the targeted culture(s) "to ensure it will be accepted or–even more critical–to ensure that it will be *loved*" (Van Boeijen and Zijlstra 2020, p. 27).

*To change a culture* suggests that the LA designer will have the intention to change a current socio-cultural value by means of a design. Yet, when considering this intention, one should be attentive to the potential challenges related to pedagogy and the organisation of education as well as cultural values that the targeted population may share. Concerning pedagogy, any LA intervention needs to be positioned in the context of the selected teaching design and educational values that are important for this societal group and also for the targeted educational institution. As stressed by Knight et al. (2014), the "relationship between learning analytics and pedagogy is important because they are both bound up in epistemology - what knowledge is" (p. 29). For example, in the context of instructionalist approaches – that assume that learning entails the transfer of knowledge from the teacher to the student or the learner (such as in the societies with high power distance) – LA's focus will be on such basic metrics as test scores, not requiring "deeper analysis of more complex artefacts, or the processes by which they were derived (p.30).

When aiming to change a culture, the designer may address the values (e.g., collaboration or trust) that are different from those that are largely accepted by the cultural group. For example, we know that some computer-supported collaborative learning activities may lead to the student's improved learning outcomes (Chen et al. 2018), but in some societies, they are not valued in the same sense as individual learning practices (Phuong-Mai et al. 2009) and related outcomes. The goal of the LA designer will be to approach the collaboration value in a way that would be easy to accept and desired by the targeted user, frequently the student or the teacher. The designer may assist in changing undesirable cultural assumptions or values, which are otherwise challenging and hard to change by for example, 1. making visible and tangible aspects of learning that were hard for people to articulate, and change, or 2. by introducing a dynamic that cultivates and also may sustain changes through for example, gamification.

*To bridge cultures* means to bring cultures together through design. In this regard, the LA designer will have an intention to elicit cooperation and respect between two or more cultures, or their selected dimensions (e.g., power distance or uncertainty avoidance) through design. One strategy can be to combine chosen values from both cultures and translate them into a new LA service.

*Bypassing culture by design* suggests that the LA designer will explicitly focus on the other aspects of design, such as the individual or universal perspective of human behaviour, the individual values of both teachers and students. That is, the LA designer should not have any culture-specific intention in mind.

Overall, a careful consideration of the design goals is critical to the successful design and implementation of LA systems. Moreover, considering the importance of understanding culture in the design and adoption of LA services, the complexity of the culture concept should not be underestimated.

One of the critical challenges in designing *culture-sensitive* LA systems and examining culture in LA research relates to an understanding of *what culture is*, given a considerable number of conceptualizations, definitions, and dimensions used to present this concept (Straub et al. 2002). Already in 1952, Kroeber and Kluckhohn identified 164 definitions of culture offered in the context of information systems research. Later, Sackman (1992) illustrated how culture can be framed in various studies as ideologies, coherent sets of beliefs, important understanding as well as basic assumptions. Further, scholars suggested that culture includes more observable, explicit artifacts such as norms and practices (e.g. Hofstede 1998) as well as symbols (e.g. Burchell et al. 1980). Schein (1985a, b) presented a three-level model of culture uncovering the more observable aspects of cultures such as artifacts and the less observable facets such as *values*. Values represent "a manifestation of culture that signifies espoused beliefs identifying what is important to a particular cultural group" (Leidner and Kayworth 2006, p. 359). In other words, these values may explain why learners or teachers behave the way they do when interacting with LA. Whereas the list of different definitions of culture presented above is not comprehensive, it provides some indication of the complex nature of culture. For the purposes of this chapter, we have chosen to approach culture from a *value-based approach,* further explained in the next section.

**10.3 A value-based approach to culture-sensitive LA design**

Even though individuals develop their cultural sensitivity through personal experiences of cultural encounters (e.g., how people collaborate or how they interpret various forms and colours), cultural sensitivity is an elastic concept that can be trained and learnt (Bennet 2004). Culture-sensitive design is seen to be beneficial for several reasons, including the following: 1. to "cross the [cultural] chasm in order to connect": cultural sensitivity in this view will enhance one's empathy and respect for the people s/he is working with (e.g., the LA stakeholders); 2. to "gain a deep understanding of the users"; cultural sensitivity makes it possible to identify what is personal, what is affected by the cultural and societal setting, and what parts of human behaviour (e.g. learning patterns) are of general character; and 3. to "be inspired to find new ideas"; the designer may use cultural differences to elicit novel ideas (Van Boejen and Zijlistra 2020, p. 30-31). Considering various definitions and conceptualizations of culture, this chapter employs a value-based approach. The values emphasized in a society may be "the most central feature of culture" (Schwartz 2006, p.139) as these values describe a shared understanding of what society views as good, right and desirable (Williams 1970). For example, if a society values success and ambition, this might be reflected in "a highly competitive economic system [...] and child-rearing practices that pressure children to achieve" (Schwartz 2006, p.139). In an educational setting, such an environment might foster competition among students as 'being better than your peers' defines a successful learner, encouraging the use of social comparison features in the design of LA dashboards (Jivet et al. 2017). We are only starting to understand how using social comparison as a reference frame is perceived and its impact on students as they use LA systems (Bennett and Folley 2021; Lim et al. 2019).

Also, in some areas of the world, some values may prevail over others, and this will influence the adoption of technologies, such as LA and artificial intelligence. For example, the Nordic Privacy Center (2020) highlights that the Nordic countries are "unique with reportedly high levels" of trust, transparency and openness as cultural values; the high levels of all these three values create a unique cultural unity, creating requirements for implementing and adopting technologies where trust, transparency, and openness are crucial.

Here, it is important to stress that both cultural values and human values may affect the needs and expectations users have towards LA as well as the implicit biases designers may pack into the implementation of LA solutions. For example, designers, based on their experience, knowledge, and views, may intentionally or unintentionally embed their own (either individual or culturally learnt) values in the LA design process, which in turn may affect the adoption of the LA service negatively. Whereas human values refer to "what is important to people in their lives, with a focus on ethics and morality" (Friedman and Hendry 2019, p.4), cultural values refer to "collective tendencies to prefer a certain course of events above another, expressed by qualifications such as good and bad, dirty and clear, ugly and beautiful" (Hofstede and Hofstede, 2005). Overall, values are characterised by the following qualities: i) values are conceptual, not physical artefacts; ii) they are not always explicit - one might act in accordance with values without being fully conscious of them; iii) values must be acted on (e.g., through the study of the student or teacher's behaviour, and iv) values consist, at their core, of "the desirable", in the sense of what is righteous (Jorgensen 2007).

## 10.4 Privacy and autonomy in LA: a value-based approach

A culturally aware LA service could reinforce or support certain values and hinder others, depending on the intentions of the designer or other factors such as stakeholders' motivation and the targeted context. For example, a LA dashboard that shares any student's progress with their peers might support the individual's learning progress but impinge on student privacy. Further, the implemented LA system may lead to improved learning outcomes for only some students. At the same time, students and teachers may trust such a system less if they were not consulted before the development was considered. While there are many values that could be considered by the LA designer (e.g., autonomy, community, fairness, equity, human dignity, inclusivity, informed consent, justice, privacy, self-efficacy, and trust), for the purposes of this chapter we have chosen to discuss two of them, namely *privacy* and *autonomy,* given the research efforts and attention given to privacy and self-regulated learning in the field of LA (Drachsler and Greller 2016; Winne 2017; Matcha et al. 2019). To complement the extensive research surrounding these two values within LA, this chapter aims to offer new insights that a cultural values-centric lens brings to these two extensively discussed topics.

### Privacy

Privacy is an elastic and complex concept and value that is associated with various definitions and interpretations. For instance, Westin (1967) defines privacy as the "desire of people to have the freedom of choice under whatever circumstances and to whatever extent they expose their attitude and behavior to others". Belanger et al. (2002) suggest that privacy refers to one's ability to control information about oneself. Further, scholars have argued that privacy represents the control of transactions between person(s) and other(s), the ultimate aim of which is to enhance an individual's autonomy and/or minimize potential risks (Dinev and Hart 2004). Overall, Smith et al. (2011) in their review of privacy in information systems research have earlier found two key definitional approaches to privacy: cognate-based and valued-based. From the cognate-based view, definitions of privacy relate to privacy as a state and privacy as control. The control-based definition has gotten into the mainstream of privacy research, "likely because it lends itself more readily to the attributes of information privacy" (Smith et al. 2011, p.995). The value-based approach sees privacy as a human right integral to society's moral value system. From the value-based perspective, privacy is overall defined as a right, and also, as a commodity.

Overall, how individuals perceive privacy as a cultural value can vary across societies. In this regard, earlier research in information systems has shown that cultural values – measured through the overall value of four cultural value indices: Power Distance, Individualism, Masculinity and Uncertainty Avoidance (Hofstede et al. 2010) – had a significant and positive influence on individuals' privacy concerns across countries (Milberg et al. 2000). This has been explained in the following way: although cultures with a high power distance index tolerate. greater levels of inequality in power, higher scores are associated with greater mistrust of more powerful groups, such as organisations or institutions. Further, cultures with a lower individualism index (i.e., collectivistic cultures), such as China, have a greater acceptance that groups, including organisations (e.g., universities), can intrude on the private life of the individual. This can directly have implications for the integration and acceptance of LA systems in some cultural contexts, but at the same time, impinges on the stakeholders' privacy. In the LA context, Hoel and Chen's (2019) findings demonstrate that there are problems using privacy concepts found in European and North-American theories to inform "privacy engineering" (i.e., "a systematic effort to embed privacy relevant legal primitives into technical and governance design", Kenny and Borkling 2002) for a cross-cultural market in the era of Big Data. That is, theories that are grounded in individualism and ideas of control of private information do not capture current global digital practice. Further, they raise the importance of a contextual and culturally aware understanding of privacy to inform "privacy engineering" without sacrificing universally shared values. In the process of "privacy engineering", the governance and work practices around it should be considered.

**Autonomy**

Etymologically, autonomy comes from Ancient Greek, with *autos* meaning self and *nomos* meaning law and defines the "ability to make your own decisions without being controlled by anyone else" (Cambridge University

Press, n.d.). Ryan and Deci (2006) argue that autonomy is a fundamental and universal human need and was one of the foundational stones on which self-determination theory was built. Autonomy is seen as a dominant value of the Western world, being central to political definitions of democracy (Blomgren 2012) or health care decision-making (Elliott 2001; Gilbar and Miola 2015). In a learning context, *autonomy* refers to "the extent to which students have choices about what to do and when and how to do it" (Schunk 2012, p.255).

Several scholars expressed doubts about the universality of autonomy across cultural contexts. For example, cross-cultural psychologists argue that autonomy is valued less by Eastern learners and question the benefits of striving for learner autonomy (Iyengar and DeVoe 2003). At the same time, other works reveal that autonomy is connected to well-being (Chirkov et al. 2003) and study success (Vansteenkiste et al. 2005) across all cultures. As a way of explaining this discrepancy, Chirkov et al. (2003) hypothesized that autonomy can be enacted differently in different cultural settings due to diverse contextual conditions. Keller (2012) proposed the same explanation and distinguished between psychological autonomy, i.e., "psychological control over intentions, wishes, and actions" (p.16) more prevalent in Western-urban environments, and action autonomy, i.e., "the responsibility to perform actions in a self-reliant way" (p.16) in rural, subsistence-based farming families.

For the purpose of this chapter, we illustrate the value of autonomy by answering the question, *Who makes decisions with LA?* We briefly discuss two instances in the context of LA design where cultural influences are worth considering: (1) a machine is making decisions instead of a human, and (2) a teacher is making decisions instead of a student. These two instances map to the two ways of using LA and educational data mining to build systems that process data and (1) make decisions automatically based on the outcomes (e.g., intelligent tutoring system, adaptive systems) or (2) report the outcome directly to the stakeholders and thus leverage human judgment (e.g., with dashboards) (Baker 2016).

Firstly, in the case of 'human vs machine', technology acceptance might impact the willingness of teachers or learners to delegate the decision-making to (intelligent) systems. The technology acceptance model (TAM) is a widely used model to understand what factors predict human acceptance or rejection of technology (Venkatesh et al. 2003), even in educational settings (Granic and Mangunic 2019) or to investigate the readiness of teachers for LA (Ali et al. 2013; Rienties et al. 2018). In this model, external variables influence the perceived usefulness and the perceived ease of use, which in turn shape behavioral intentions and lead to the actual use of the system (Venkatesh et al. 2003). TAM was developed in the US and was widely used across cultures, but there is some evidence that TAM might not hold in all cultures (Srite 2006), as certain cultural orientations "nullify the effects of Perceived Ease of Use and/or Perceived Usefulness" (McCoy et al. 2007, p. 81).

Undoubtedly, *trust* is another factor that influences learners' and also teachers' willingness to allow systems, AI educational technologies, in particular, to make decisions for them (Nazaretsky et al. 2022). As cultural norms and values shape trust in human relationships (Doney et al. 1998), they also hold human trust and attitudes towards automation (Chien et al. 2016). Thus, next to digital literacy, culture could also play a role in LA design decisions around who should retain the power of decision, who would take responsibility when the system makes a mistake (Santoni de Sio and Mecacci 2021), and how the system communicates the educational values the school is supposed to teach.

Secondly, one can look at the interplay between cultural values, the teacher-student relationship, and expectations around LA. Over the past years, there has been a strong focus on supporting self-regulated learning through LA (Winne 2017; Viberg et al. 2020), enhancing and developing autonomy in students, and equipping them with skills to become masters of their own learning. Again, this focus has been shaped in the Western context from a Western perspective. As previously mentioned, research suggests that autonomy might benefit all students, regardless of their cultural background. Initial work exploring teachers' expectations across continents has shown contrasting outcomes. While in Europe teachers expect LA to enable decision-making on the student side (Kollom et al. 2021), LATAM teachers find more value in LA tools supporting teacher decision-making (Pontual et al. 2022). LA – as an educational research tool – can be designed to provide computational proxies for student levels of SRL (e.g. Viberg et al. 2020). That is, it could be used to help investigate if there are cultural differences in student SRL. Furthermore, if LA can then be used to support students, e.g., through the provision of feedback on SRL, then one can study the differential effects of doing so with different student cultures. This in turn may connect with feedback literacy, i.e., learner disposition and skills of actively seeking out and engaging with feedback (see e.g., Lim et al. 2021). In sum, these are just a few examples that showcase the importance of considering cultural differences when designing and implementing LA systems across countries.

## 10.5 Future research directions

Based on the argumentation and a scoping review of the literature presented in this chapter, there are several future research directions to be considered by the LA community.

Grounded in the extensive research on culture in the established field of information systems research (for an overview, see Leidner and Kayworth 2006), LA scholars need to look into the question of how culture influences stakeholders' requirements for LA systems' design. Here, it is important to consider different levels of culture, including national and organizational ones (e.g., educational institutions often have their own cultures). One challenge relates to the assumption that all individuals within a selected cultural unit will respond in a consistent way based on the group's cultural values; this view limits individual differences that may be found within a particular cultural unit (e.g., a school) that may lead to different behavioural outcomes. To better understand such differences, the application of person-centric approaches (Hickendorff et al. 2018) offers a solution.

Further, when conducting related cross-cultural studies (e.g., evaluating the use of a selected LA tool in the selected contexts) LA researchers need to address three types of methodological bias – found in other information systems-culture research (Leidner and Kayworth 2006). First, there is *construct bias*, suggesting that a given concept or cultural value (e.g., privacy) is not viewed similarly across contexts. Second, there is *method bias*, i.e., when study participants across cultures and countries do not respond similarly to measurement scales due to factors linked to demographics or the administration of the instrument. And finally, there is *item bias* that derives from the poor translation of the (survey) instrument.

## 10.6 Conclusion

In this chapter, we argued how culture can play a role in the design of LA and proposed that addressing factors such as stakeholders' cultural values can influence the successful adoption of LA at scale. We stressed the importance of the cultural setting, and the range of intentions designers can have when dealing with culture (affirming, attuning, changing, abridging, bypassing) as this helps to set the tone for the envisioned LA system. We then specifically looked at two cultural values, i.e., privacy and autonomy, to exemplify how such values might affect the requirements for and the design of LA systems.

Based on the argumentation provided above, we outline several design implications for culturally aware LA, mainly aimed at learning (analytics) designers. Nonetheless, these suggestions could also be followed by teachers adopting LA solutions in their classes.

*First*, designers who aim to develop culturally-aware LA solutions need to start with the definition of culture and values. This can be achieved through the examination of existing understandings of culture and values, both in research and practice. Such definitions can be also offered by stakeholders (students and teachers).

*Second*, when designing culturally aware LA systems, designers need to consider and differentiate between individual and cultural *educational* values. Although a group's culture might shape the values of its members, personal values among the individuals of the same group vary.

*Third*, there is a need to keep the design intention in mind, for example, whether the suggested LA system aims to affirm cultures or bridge them. This is important to consider from the very start of the design process. This decision can be informed by key stakeholders directly, which relates to the need to use participatory design and co-design approaches for a more likely adoption of LA systems (see e.g. Sarmiento and Wise 2022).

*Fourth*, designers could consider using existing culturally aware and value-sensitive design methods, originating in other disciplines, such as human-computer interaction in the LA design process. Such methods include direct and indirect stakeholder analysis, stakeholder tokens, value-source analysis, value scenario, and value-oriented semi-structured interviews (Friedman and Hendry 2019).

And finally*, fifth*, one should pay attention to the fact that culture and values (such as privacy) are elastic concepts and may change over time. This is important to keep in mind to be able to adapt to the stakeholders' views, needs and preferences concerning the implementation of LA systems. In this regard, the application of for example, Design Science Research methodology (DSR; Hevner et al. 2004; March and Smith 1995; Vaishnavi and Kuechler 2008), also valuable in education settings (Laurillard 2012) can be helpful.

To conclude, we would like to stress the importance of looking at culture in a global educational landscape that is ever increasing without reducing learners to their culture. Our intention is not to suggest that individuals can be prescribed a certain type of (learning) analytics based on their culture, but rather to cultivate an awareness of the influence that cultural values might have on the perceptions and preferences of both learners and teachers.